\documentclass[journal,onecolumn]{IEEEtran}  

\usepackage{graphics} 
\usepackage{graphicx}
\usepackage[table,xcdraw]{xcolor}
\usepackage{multirow}
\usepackage{amsmath} 
\usepackage{amssymb}  
\usepackage{hyperref}
\usepackage{array}

\begin{document}

\title{ 
Objective Multi-variable Classification and Inference of Biological Neuronal Networks
}

\author{Michael Taynnan Barros, Harun Siljak, Peter Mullen,\\ Constantinos Papadias, Jari Hyttinen and Nicola Marchetti
\thanks{M. T. Barros and Jari Hyttinen are with the CBIG/BioMediTech, Faculty of Medicine and Health Technology, Tampere University, Finland. email: michael.barros@tuni.fi}
\thanks{H. Siljak, P. Mullen, and N. Marchetti are with Trinity College Dublin, Ireland.}
\thanks{C. Papadias is with the American College of Greece, Greece.}}

\markboth{Version to ArXiv}%
{Us \MakeLowercase{\textit{et al.}}Capacity of Exosomes Com. Sys.}

\maketitle

\begin{abstract}
Classification of biological neuron types and networks poses challenges to the full understanding of the brain's organisation and functioning. In this paper, we develop a novel objective classification model of biological neuronal types and networks based on the communication metrics of neurons. This presents advantages against the existing approaches since the mutual information or the delay between neurons obtained from spike trains are more abundant data compare to conventional morphological data.
We firstly designed two open-access supporting computational platforms of various neuronal circuits from the Blue Brain Project realistic models, named Neurpy and Neurgen. Then we investigate how the concept of network tomography could be
achieved with cortical neuronal circuits for morphological, topological and electrical classification of neurons. We extract the simulated data to many different classifiers (including SVM, Decision Trees, Random Forest, and Artificial Neuron Networks) classifying the specific cell type (and sub-group types) achieving accuracies of up to 70\%. Inference of biological network structures using network tomography reached up to 65\% of accuracy. We also analysed recall, precision and F1score of the classification of five layers, 25 cell m-types, and 14 cell e-types.
Our research not only contributes to existing classification efforts but sets the road-map for future usage of cellular-scaled brain-machine interfaces for in-vivo objective classification of neurons as a sensing mechanism of the brain's structure.

\end{abstract}



\section{Introduction}

The detailed characterisation of the human brain has recently gathered investment of many countries, in both academia and industry, to not only create a digital reconstruction of the brain but also to move towards a complete understanding of its mysteries \cite{reconSim}. The key components of these efforts are the many morphological and electrical types of neurons, which impact on the linkage between cellular intrinsic properties to the whole brain's functioning, behaviour and further pathology characterisation. Without solving this issue, the neuroscience community will remain to base their analysis on the unverified subjective classification of neurons, which is far from a precise approach \cite{kanari2019objective,vasques2016morphological}. Besides, this type of information is crucial for not only neuron type target drugs but to precise micro-scale brain-machine interfaces that can interact at the cellular level to uncover small-scale information in the brain \cite{barros2019topology,balasubramaniam2018wireless}.

The recurring issue of classifying morphological neuron structures, despite the many anatomical studies, is about the lack of consistent results by methods of visual inspection of brain slices driven by experts on the field \cite{defelipe2013new}. This issue can be solved with objective methods that use rigorous analysis to introduce less bias in the classification method. More recently, proposed methods have concentrated on supervised machine learning techniques, where a training dataset is created based on metrics that characterise the cellular morphological structure \cite{kanari2019objective,vasques2016morphological}. This is a more formal technique compared to visualisation techniques counterpart, which depends on a trained machine learning solution. Even though these new approaches are exciting, we hypothesise that based on the relationship of cellular morphology and activity \cite{deitcher2017comprehensive}, we can perform reliable objective morphological classification based on cellular activity and communication alone. In this way, we would provide advantages to the existing approaches, where activity and communication are more accessible and abundant data than the existing counterpart. Our approach can also be used as a complementary approach to the existing efforts in neuron type identification even though we do not explore this idea in our paper. 

In this paper, we propose a new objective morphological classification based on the activity and communication of neurons alone by \textit{i) characterising their information transfer using classical Shannon mutual information theory and spike delay estimation.} The signal propagation inside the circuits is measured by the level of information communicated between cells.  \textit{ii) developing new simulation support libraries for integrating large cortical circuit formation using the Python language to link the validated computational models of the Blue Brain Project and the NEURON tool, called Neurpy and NeurGen.} Our neuronal computational framework is depicted in Fig. \ref{fig:nrpySimFw}. Using those we can generate neuron signal data as they exist in a realistic cortical network, to measure the effects of cell-to-cell connection parameters on the inter-layer neural information transfer. We, therefore, were able to characterise the response of individual neural cells in a finite dimension space using the Linear-Nonlinear-Poisson cascade model. \textit{iii) developing a classification system to predict cell type (and sub-group type) and networks based on network tomography method}.
We investigate the classification and inference of neuronal cells and circuits, especially at an inter-cortical layer scale and with a larger variety of cell-types, progressing the field towards the classification of more complex cortical circuits. We build several supervised classification algorithms using features from our information and communication analysis based on various forms of cortical networks, from known networks to existing ones discussed in \cite{bbpTop}. We performed feature engineering to reduce the data dimensionality and eliminate biases by filtering the training data, which we either applied to the neural network and/or decision tree classification models leading to superior performance. Finally, we reconstructed the cell-types in a known topology based on endpoint measurements taken around the network.

\subsection{Contributions}
Contributions from our research are as follows:

\begin{itemize}
    \item \textbf{A simulation support library referred to as \emph{Neurpy}:} that allows for the rapid generation, construction and, simulation of cortical networks. This library is written in Python and uses the NEURON-backend API while accepting XML-based network descriptors to define the cortical circuit.
    \item \textbf{A network-generation script termed \emph{NeurGen}}, which uses statistical information of neuronal pathways to generate large quantities of unique networks. The generated networks can be either random in shape or can be specified to fit a given topology template. The generated network descriptor is in a format which allows seamless integration with the \emph{Neurpy} library.
    \item \textbf{The usage of discrete-memoryless mutual information between two cells for the characterisation of the intercellular information transfer:} We describe the process of train-discretisation for the conversion from a set of analogue voltage measurements into a binary sequence. This binary sequence is then used to calculate the discrete-memoryless mutual information between two cells. We also describe the process of simple delay-estimation using cross-correlation of the network measurements, selecting peaks in the correlated result to predict the synaptic delay, which we find to have promising results.
    \item \textbf{Classification and inference of biological neural network morphology and topology based on information network tomography:} We then propose and implement a method of cell-characterisation through filter coefficient estimation using the pseudo-inverse method for resolving linear equations, similar to existing cell-description models such as the \emph{Linear-Nonlinear-Poisson} cascade model. Using this cell-characterisation we describe the process of the training of many cell-type classification models to predict cell sub-groups from input-output voltage measurements. Finally, we describe the process of topology-reconstruction using the supervised classification models to estimate the cell-type in a 4-leaf star topology. Three different classification algorithms are used independently to estimate the neuron layer, its electrical type and its morphological type respectively for each element in the network.
\end{itemize}

\subsection{Related Work}

Investigation of the information transfer in neuron communications is an active topic in the research area of molecular communications to further understand biological signal propagation \cite{akan2016fundamentals}. Even though researchers have concentrated on the molecular synaptic transfer \cite{ramezani2018impacts}, or single-cell models \cite{balevi2013physical}, works such as \cite{barros2018capacity,veletic2016peer} progressed towards neuron population scenarios to analyse more impacting information transfer in the brain.
Prior work investigating the mutual information in neural communications presented in \cite{spikeTrainInfo} has expressed the entropy of a source through the time-binning of the spike train, and defining the set of possible symbols as the possible neural response within the period. Other methods were based on type-rich network structures but suggested that network connections are more relevant to the information transfer inside a network than the main cellular differences between types \cite{barros2018capacity}. The main issue presented in their research is that a large data set is required to converge on a true entropy value. A possible workaround that was presented is to instead calculate a lower and upper bound on the entropy value which is robust regardless of dataset length. From their definition of the entropy boundaries, it was found that the information rate (i.e. the mutual information) of the system increased with increasing entropy, as expected. We aim to use their proposed model to measure the communication between neurons.

Works like \cite{defelipe2013new,kanari2019objective,vasques2016morphological} have provided many contributions towards the objective classification of neurons. They are interested in the morphological structures and classification based on that. The major criticism is that morphological training data is difficult to accumulate, and this limits the creation of adaptable classification techniques \cite{glaser2019roles}. Morphological data also limits the usage of existing approaches to more complex structures, including cortical microcircuits. Our objective is to make objective neuron classification expand in different ways, i) we aim to make it more reliable ii) make it useful for other higher complex neuron structures and networks and iii) provide a method that is more accessible for other researchers in the area or those dependent on these classification methods for other tasks.

Recently, Barros et al \cite{barros2019topology} investigated the classification of cortical network parameters given simulated endpoint data using a theory called \textit{Network Tomography}. 
The overall accuracy of this classification model was about 83\%, with the model performing well for four different combinations of four pre-synaptic groups.
As well as cell-group classification, a topology-discovery classifier was also investigated in that work. 
The topologies investigated were relatively small and the classifier achieved accuracy as high as 99.37\% when classifying the 2-leaf and 4-leaf topologies through the use of decision trees using the same cell types.
While the research conducted focused mainly on a relatively small number of same type neurons, we expand on this by investigating similar types of classification with more complex sets of classes towards more realistic cortical structures, such as with the full set of cell-possibilities, which was completely abstracted in the work of \cite{barros2019topology}.


\section{Methods}

\begin{figure*}[ht]
    \centering
    \includegraphics[width=\textwidth]{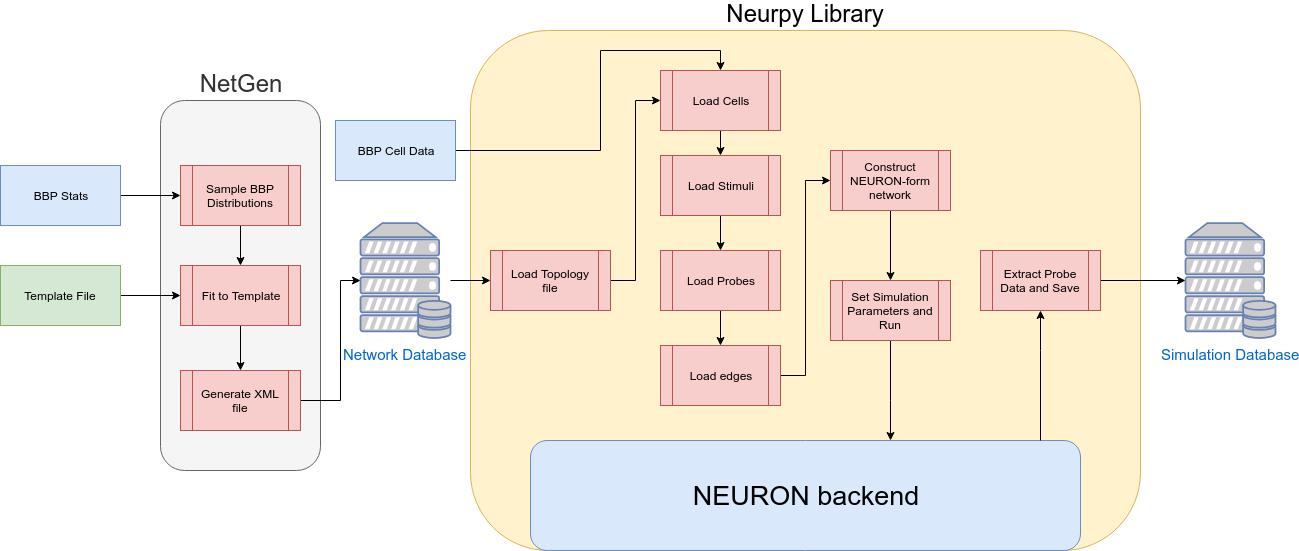}
    \caption{Structure of experimental environment: network (topology) generation, cell information input and simulation}
    \label{fig:nrpySimFw}
\end{figure*}

\subsection{Cortical Networks Characterisation} \label{sec:nrnBack}

In this work, we focus on the neurons of the somatosensory cortex of a juvenile rat. 
The neocortex is typically treated as being a tightly-packed system of vertical structures (cortical columns) and classified into 6 interconnected horizontal layers, referred to as L1-L6 (with layers 2 and 3 often grouped and referred to as L2/3). Neurons can be classified by morphological-type (m-type) which defines the physical shape and layout of the neuron, and by electrical type (e-type) which defines the electrical characteristics of the cell. The neurons in this investigation are based on data retrieved from a single cortical column, where each neuron is pre-defined by layer, m-type, and e-type.


\subsubsection{Cell Characterisation} We focus our modelling on the characterisation of the response of the many neurons m-types to a given stimulus signal. This response is an \emph{all-or-none} spike train response. We do not explore the description biophysical model of neurons in the paper since we believe that information is covered by the NEURON framework and the Blue Brain Project. Rather, we focus on the analysis of each neuron as a high-level system.
As the release of neurotransmitters in the synapse is a response to a spike event rather than to the small variations in resting potential, the response of a postsynaptic cell is therefore also a function of the spike events. Analytically, the spike events are treated in a probabilistic fashion (i.e. the probability of a spike occurring) to which a Poisson process fits well \cite{herfurth2019quantifying}. The Poisson process describes a model for the estimated time between events, and so when applied in this domain to estimate the time between voltage spikes, the intercellular signals are seen as a \emph{Poisson Spike Train}. This has several useful implications for the analysis of these intercellular signals in that the probabilistic modelling of the signal allows for the application of the Poisson model in information and communication domains that rely on signal probabilities, such as mutual information and delay. 

Another implication of this spike-triggered-event property is that the cell can be described as a time-varying process responding to (and with) an impulse train. One such model commonly used is the \emph{Linear-Nonlinear-Poisson cascade} (LNP) model discussed in \cite{lnp} where the response of a cell is described by three serial components: a linear filter, a non-linear transform, and a Poisson spike-generating process. In the LNP model, the dimensionality reduction occurs in the linear-filter component (the non-linearity and Poisson component deal mainly with the production of a spike train which is not required here). While this model generally deals with a spatio-temporal input (i.e. an external screen), we can adapt the concept to use the output spike-train of another neuron as input. As the spike-train is essentially an impulse-train, the resulting linear filter that best characterises the neuron can be estimated as the impulse response of a time-varying system taking an impulse train as input and generating a continuous voltage-time output. In our case, a finite impulse response (FIR) filter model was used. 
By measuring the input and output spike trains of a single cell, an estimation can be determined of the k-order FIR filter that best characterises the neuron's impulse response \cite{sysIdA}, \cite{sysIdB}. The theoretical principle here is, given the linear system 

\begin{equation}
    \label{eq:firLinSys}
    y = x \star h,
\end{equation}

\noindent one needs to estimate the value of $h$ to a limited order that best produces the output sequence $y$ from input sequence $x$ (where $\star$ denotes the convolution operation.) If we take this to be representative of a finite-impulse response (FIR) filter, we can describe the system as $\vec{y} = \boldsymbol{X}\vec{h}$ where $\vec{h}$ is a vector of the coefficients of a $K$-order FIR filter, $\vec{y}$ is the vector of $N$ output measurements, and $\boldsymbol{X}$ is an $N\times K$ matrix representing the time-shifted window of input measurements, through $N$ time-shifts and a window size of $K$. This system is not invertible to solve for $\vec{h}$ where $\boldsymbol{X}$ is not square, however, it is possible to solve for $h$ under "general conditions" through the use of the \emph{pseudoinverse} of $\boldsymbol{X}$. This can be more specifically defined by the Moore-Penrose pseudoinverse of $\boldsymbol{X}$, so $\vec{h}$ is given by

\begin{equation}
    \label{eq:estFilter}
    \vec{h} = (\boldsymbol{X}^{T}\boldsymbol{X})^{-1}\boldsymbol{X}^{T}\vec{y},
\end{equation}

\noindent where $\vec{h}$ is the estimated filter, $\boldsymbol{X}$ is the equivalent time-delayed input matrix, and $\vec{y}$ is the output signal vector.

\subsubsection{Train Discretisation and Probability Analysis}
The first step in the probabilistic analysis is to discretise the voltage-time spike trains into a binary series, similar to the process followed in \cite{spikeTrainInfo}. In our case, the process undertaken was as follows:

\begin{enumerate}
    \item Threshold the spike-train - values above the voltage threshold $V_{thr}$ are set to 1, all other values set to 0
    \item Subdivide the thresholded train into many windows of length $L_{w}$
    \item Check for the presence of a spike within the window, set the window-value to 1 if found, 0 otherwise
    \item Convert the window-interval values into a binary sequence
\end{enumerate}

The binary output sequence of this process represents the symbol of the system; in this case, our symbol is 1 bit and so our symbol set can be described as $S=\{s_{1}=0, s_{2}=1\}$. Here we can calculate the binary sequence from the input, the head-cell (which we can represent as a random variable $X$) as well as from the output, the tail-cell (represented as random variable $Y$). A head-cell is the neuron with the highest connection degree inside the network, and the tail-cell the one with the lowest. With this in mind, we can use the entropy model of the head-cell and mutual information of the cortical link, which is presented in the following. First, we calculate the probability mass function (PMF) of both X and Y (that is, $P(X)$ and $P(Y)$) from their respective binary sequences. Then we apply various forms of Bayes' rule, along with further analysis of the two binary sequences, to obtain the joint PMF ($P(X,Y)$) and the conditional PMF ($P(X|Y)$) for the calculation of the mutual information between a head-cell and a tail-cell.

\subsubsection{Cortical Topologies} There are two main topologies investigated in this study: 2-cell topology, and 4-leaf star topology. The 2-cell topology consists of two cells linked by a synaptic connection whose parameters vary from network to network. It is a simple topology use as a starting point in this study mainly to analyse the link between two neurons rather than to investigate the various effects of higher-complexity neuronal layouts. The 4-leaf star topology is only a slightly more complex network, but enough to allow for richer behaviour. The use of this network intends to analyse the application of the results found from the study of the 2-cell networks in more complex topologies to investigate what effects if any, the topology may have on performance. It is important to note here that the direction of the synaptic connections is that of an "out-hub", i.e. the central node (acting as the hub) is delivering voltage spikes to the surrounding leaves. The main reason for starting with these topologies' size is that they mimic the small-world topology, which is believed to be the underlying structure of complex networks in the brain \cite{bassett2017small}.

\subsection{Information Theoretic-based Network Tomography based on Spike Delay Estimation} \label{sec:netTomInfoBack}

Network tomography is a branch of information and communications theory that deals with the inference of internal network properties from a finite number of endpoint measurements. This concept can be applied in many specific applications, however, it is generally used in internet systems to determine link-loss and link-delay characteristics as discussed in \cite{intTom}. Analytically speaking, the problem of broad inference of the entire network can be approximated by describing the measurements as a linear model \cite{intTom} given by:

\begin{equation}
    \label{generalNetTom}
    \vec{y}=\boldsymbol{A}\vec{\theta}+\epsilon,
\end{equation}

\noindent where $\vec{y}$ is a vector of measurements, $\boldsymbol{A}$ is a routing matrix representing the network node connectivity, $\vec{\theta}$ is a vector of link parameters (delay, loss etc.), and $\epsilon$ is a noise vector. The routing network is typically a binary matrix with the $(i,j)^{th}$ element being 1 to represent a connection between the $i^{th}$ node and the $j^{th}$ node, and 0 representing no connection between the two nodes. The network inference, in this case, would be estimating the vector $\vec{\theta}$ given some endpoint measurements $\vec{y}$ and knowledge of the networking routing matrix, along with some distribution for the noise parameter $\epsilon$ (Gaussian, Poisson, etc.). For large networks, this poses a problem in the computational solution for the linear model, as the dimensionality of $\boldsymbol{A}$ can grow prohibitively large for larger networks. 
It is worth noting that the usual networks subject to tomography (e.g. internet) are bi-directional. In cortical circuits, the cell-to-cell links are mostly unidirectional with signals propagating in a single direction. This has many implications for the routing matrix $\boldsymbol{A}$, namely that the presence of a 1 for element $\boldsymbol{A}_{i,j}$ implies a high probability of a 0 for element $\boldsymbol{A}_{j,i}$.\\
We also observe our networks from the \emph{information theoretic} standpoint. The information theory of discrete systems deals largely with symbols. As there is a close link between information theory and probability, a symbol can be thought of being similar to a discrete random variable, where a source may produce a symbol at every event from a set of possible symbol values. The information contained in a given symbol is defined by $I(s_{k}) = -\log_{2}(p_{k})$ where $s_{k}$ belongs to the symbol set $\{s_{1},\ldots,s_{N}\}$ and $p_{k}$ is the probability of the corresponding symbol occurring. By taking the \emph{average} entropy of each symbol in a given set (weighted by symbol probability), we obtain the \emph{entropy} of the set, defined as $H(X) = -\sum_{k=1}^{N}p(s_{k})\log_{2}(p(s_{k}))$ where $H(X)$ is the expected uncertainty of event source $X$. 
Another useful metric in information theory is that of \emph{conditional entropy} or the entropy of some random variable given the knowledge of another random variable. Given the random variable $X$, the entropy of $X$ given a particular value of another random variable $Y=y_{k}$ is given by

\begin{equation}
\begin{aligned}
    \label{eq:condEnt}
    & H(X|Y) = \sum_{k=1}^{N}H(X|Y=y_{k})p(y_{k}) \\
    & = -\sum_{k=1}^{N}\sum_{j=1}^{N}p(x_{j},y_{k})\log_{2}(p(x_{j}|y_{k})),
\end{aligned}
\end{equation}

\noindent where $H(X|Y)$ is the conditional entropy of $X$ given $Y$, $p(x_{j},y_{k})$ is the joint probability of symbols $x_{j}$ and $y_{k}$, and $p(x_{j}|y_{k})$ is the conditional probability of the same symbols. Given this definition of the conditional entropy, we can obtain an expression for the reduction in entropy of $X$, given some observed event $Y$. This is defined by

\begin{equation}
    \label{eq:mutualInf}
    I(X;Y) = H(X) - H(X|Y),
\end{equation}

where $I(X;Y)$ is referred to as the \emph{mutual information} of $X$ and $Y$. 


\subsubsection{Delay Estimation}


Following the collection of the simulation data, our analysis concentrates on the estimation of the link delay through the use of the cross-correlation between the two measured voltage plots, $x_n$ for the head-cell and $y_n$ for the tail-cell where $n \in \Re^+$. This was implemented by the cross-correlation of these two series. First, we define the true cross-correlation of $x_n$ and $y_n$ as

\begin{equation}
    R_{x,y}(m) = \mbox{E}[x_{n+m}y^*_n] = \mbox{E}[x^*_n y_{n+m}],
\end{equation}

\noindent where the asterisk denotes the complex conjugation, and $\mbox{E}[.]$ denotes the expected value operator. Since $n$ is an infinite-length random variable, we can only estimate a real cross-correlation value. Therefore, we may define

\begin{equation} \label{eq:cross}
\hat{R} =
\begin{cases}
\sum^{N-m-1}_{n=0} x_{n+m}y^*_n, m \geq 0\\
\hat{R}^*_{x,y}(-m), m < 0,
\end{cases}
\end{equation}

\noindent which will lead to a final cross correlation of $\hat{R}_{x,y}^* (m-N)$ with $m=1,2,...,2N-1$.

Next, we find the closest positive peak in the cross-correlation using local maxima theorem in a series. We define the local maxima as the portions of the plot where the first derivative is 0 and the second derivative is negative (concave-down). By assuming that the first local maxima represent the delay, we can obtain a simple estimate of the link delay. 

Conceptually, a spike in the head-cell should result in a spike in the tail-cell separated in time roughly by the delay as the spike crosses the synaptic connection. As a result, we should find a peak in the cross-correlation between the voltage measurements of the two cells as the time shift approaches the delay since at this time-shift the signals should be relatively well correlated.ß
While this approach is quite basic and may not be overly accurate, it is a first step in expressing the characteristic function of the delay in the neuronal link, which is an important concept when applying network tomography to estimate link-level delay, as discussed in \cite{netTomFour}.

\subsection{Neuronal Computational Framework}
\label{sec:simFrame}
Here we discuss the design and implementation of the simulation framework used to generate, construct, simulate, and analyse the neuronal circuits. 

\subsubsection{The NEURON tool}
The NEURON tool is a simulation framework developed by researchers at Yale University for the \emph{in-silico} analysis of individual and networked neurons. The framework provides some tools for the construction, simulation, and recording of individual cell-models from biological principle building-blocks as well as networks built from individual neuronal models (see \cite{neuronSimEnv} for more details). 

\subsubsection{Cell-Data Source}
The cell data used by NEURON to load and simulate individual cells were provided by the Blue Brain Project (BBP) through the Neocortical Micro-Circuit (NMC) portal available at \cite{nmcPortal}. This data is supplied in a format that is compatible with the NEURON framework, including sample script files for initiating and simulating single-cell networks. Each cell supplied is categorised by the layer, m-type, e-type, and the variant number (multiple variants may exist for the same layer;m-type;e-type group). For example, the cell titled "L1\_DAC\_bNAC219\_1" represents a layer 1 cell of DAC m-type and bNAC e-type. For each cell, a number of data-files are supplied. One of the main files is the morphology descriptor. This file contains the complete description of the entire morphology of the cell, formatted as the NEURON-compatible section-segment hierarchy which can be loaded quickly into the framework. The supplied data describes the cell's biophysical properties. These properties represent the specific electrical characteristics of various sections within the neuron and are vital to the accurate simulation of the cell. 

\subsubsection{Python Support Library}
While the sample files given by the BBP are extremely useful and informative, they are limited in that it is not inherently easy to connect multiple cells to form a multi-cellular network. As well as this, the use of the default GUI is not a scalable approach when generating a large set of training data. For this reason, we developed a Python library to address these issues. The library, referred to as \emph{Neurpy} and available at \cite{neurpyGit} uses the NEURON-Python API to more easily construct simulations of any network form, while also allowing for either GUI interfacing (for interactive/debug purposes) as well as a more lightweight "headless" interface where the simulations are run entirely through the Python code without any need for user interaction. A network generator script named \emph{NeurGen} was constructed in Python. NeurGen functions by loading the JSON-formatted statistical data on the physiology and anatomy of each cellular pathway, building an internal database of statistical distributions for each of the connective parameters, and then generating a network which conforms to the given distributions. For this study, we developed the option of using topology templates as well. The template defines the overall network shape (number of cells, inter-cell connections, the stimuli, and the probes) without requiring any specific data to be specified about any of the network components. Cell-type constraints are also accepted in the template, which limits the set of possible neural cells that can be selected for a given position in the topology. The complete code release is available at \cite{codeGit}.

\subsubsection{Overview of Simulation Framework}

At this point, all the components of the simulation framework have been defined. An overview of the entire process is shown in Figure \ref{fig:nrpySimFw}, with the general flow of the process being from left to right. On the left, we begin with the network generator, where we feed the statistical data on the cell-to-cell pathways from the BBP along with a topology template into NeurGen. Within NeurGen we can then generate a large number of unique individual networks, which we store as the ”Network Database”, a collection of XML files describing each network. After the database of networks has been created, we can begin feeding this into the Neurpy library, file by file. For each network file, Neurpy loads the specified cell model data from the BBP, loads and connects the required stimuli, loads and connects the required measuring probes, and finally connects the cells to form the network. Following this, the session simulation parameters are set (timestep, simulation length etc.) and Neurpy instructs the NEURON backend to begin the simulation. After the simulation has completed, all data is extracted from the probe vectors, converted and concatenated in a Python-native format, before being written to the disk. By repeating this for all network files, we create the "Simulation Database" which contains the voltage-vs-time data for each probe in each simulation, which we can then use for analysis.

\subsubsection{Simulation Dataset}
The cortical networks for the study of the neuronal communication network parameters were generated using the NeurGen script described above. A simple 2-cell topology template was passed to the generator which defined a network of 2 cells (with no constraint on cell type), a stimulus on the head-cell, and a probe on each soma. As this portion of the investigation deals with the communication network parameters between any 2 cells in the cortical circuits, there was no need for a topology more complex than 2 cells. Consideration of more than 2 cell connections could lead to multi-path problems, which are not investigated in this paper. The simulator script was then modified slightly so that after the loading of each topology (but before the running of the simulation) some parameters of the communication link were forcibly varied. These parameters included the link delay and multiplicative gain (weight), the distance between the nodes, as well as the interval, delay, weight, and symbol probability of the stimulus. These variations were output to a separate "metadata" file for each simulation such that they could be used in the analysis to find correlations in the data. After this, the network was simulated with a simulation length of 1000ms.

\subsection{Cell Classification}
As previously mentioned, the Linear-Nonlinear-Poisson (LNP) cascade model can be used to model and characterise the response of a neural cell. Generally, this is computed using the spike-triggered-average (STA) of the stimulus sequence; however, this approach is more appropriate when dealing with external stimuli such as a spatio-temporal screen, discussed in \cite{lnpInBrain}. For this reason, the FIR-filter estimation method described in Equation \ref{eq:estFilter} was used to reduce the dimensionality of the measured features. 

To investigate the objective classification of biological neuronal structures and to train the support vector machine (SVM) and decision tree classifiers, Matlab was used. Matlab was used due to its community-backed \emph{Classification Learner} tool. This is a tool that takes a dataset as input, requests the specification of the features and classes in the dataset, and then quickly trains a number of different classifiers against the data, reporting the individual classifier accuracy, confusion matrix, and receiver-operating-characteristic (ROC) curve. This tool is therefore very useful for quickly getting an idea of how the different forms of classifiers are dealing with the given features and classes. We also trained random forest and artificial neural network classifiers using RapidMiner, which offers a powerful environment for developing data processing and machine learning models besides those included in the Matlab tool.

We have previously discussed how each of the cells in the dataset has been classified by the BBP; that is, each cell has an associated layer, m-type, and e-type. As there are over 1000 individual cell models, it is unlikely that a classifier will be capable of gaining any form of functional accuracy in directly classifying the exact cell type. For this reason, the classification of a given cell was broken down into the classification of each of its constituent components, i.e. 3 separate classifiers were trained to estimate the overall cell type. The first classifier estimates the layer to which the cell belongs based on the filter coefficients extracted through equation \ref{eq:estFilter}. The second classifier estimates e-type based on the filter coefficients as well as the output of the first classifier (the layer estimation). The final classifier estimates m-type based on the filter coefficients, the estimated layer, and the estimated e-type. In this format, each classifier is only estimated against 5, 11, and 24 classes respectively. The other benefit of this approach is that the output of one classifier can be fed into the next, which allows it to take into account the associated probability of one class leading to another.

\section{Results}
\label{sec:res}

\subsubsection{Delay Estimation}

\begin{figure}[ht]
    \centering
    \includegraphics[width=0.48\textwidth]{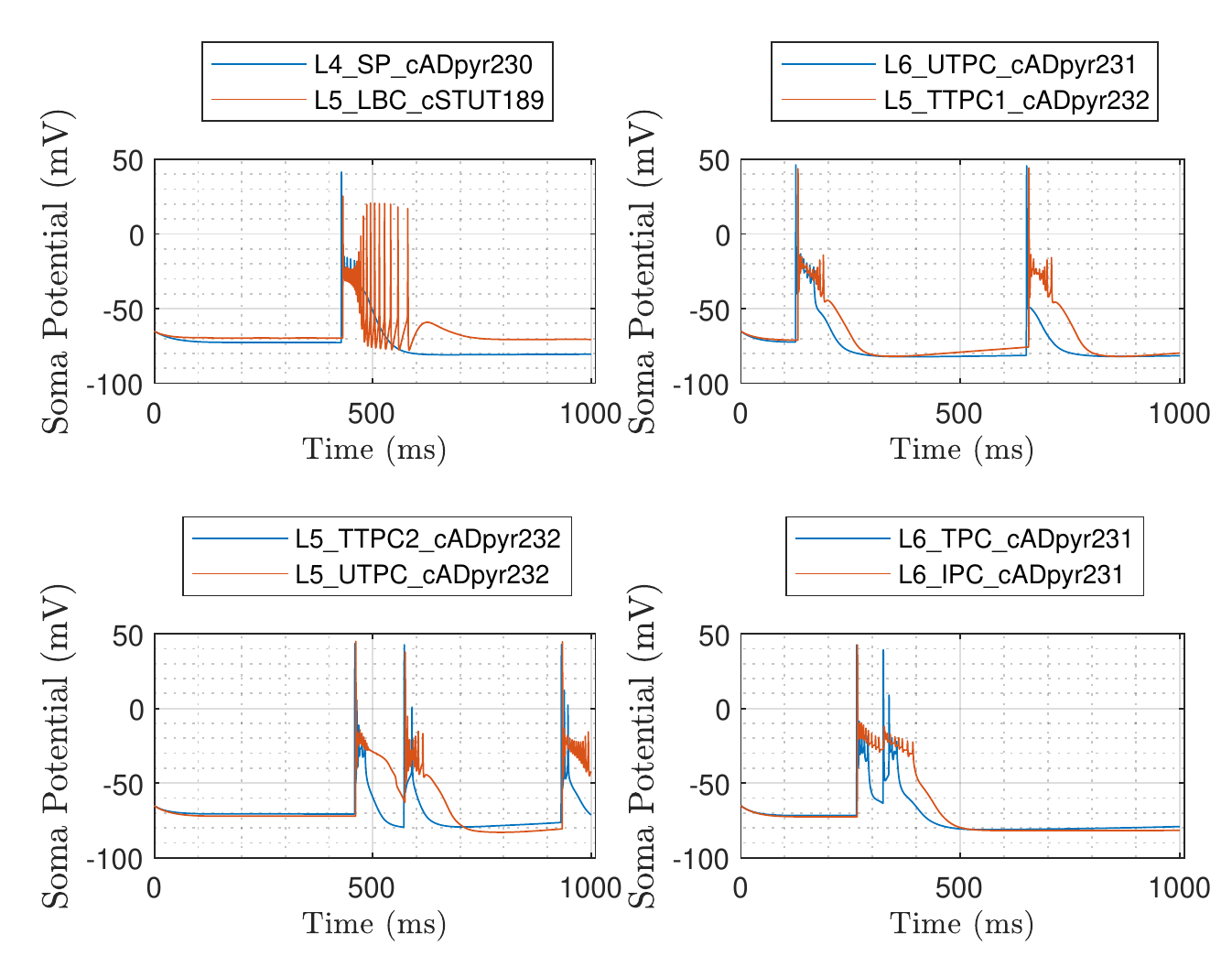}
    \caption{Comparison of 2-cell network outputs. head-cell in blue, tail-cell in orange. Cell-types shown in each subplot's legend.}
    \label{fig:sample2CellPlots}
\end{figure}

The analysis of the communication network details of a synaptic connection was done through the simulation of 10,000 simple 2-cell networks. This results in the production of sets of voltage data for each simulation, one from the head-cell and one from the tail-cell. As the output of one cell is acting as the stimulus to the other, the spike trains from the simulations tend to be correlated. The output plots from a few of these simulations are shown in Fig. \ref{fig:sample2CellPlots}. In these sample plots, the source-destination spike correlation is clear, as a spike in the pre-synaptic cell often results in a spike in the post-synaptic cell. In addition, the plots indicate a slight delay between the source and event spikes.

\begin{figure}[ht]
    \centering
    \includegraphics[width=0.48\textwidth]{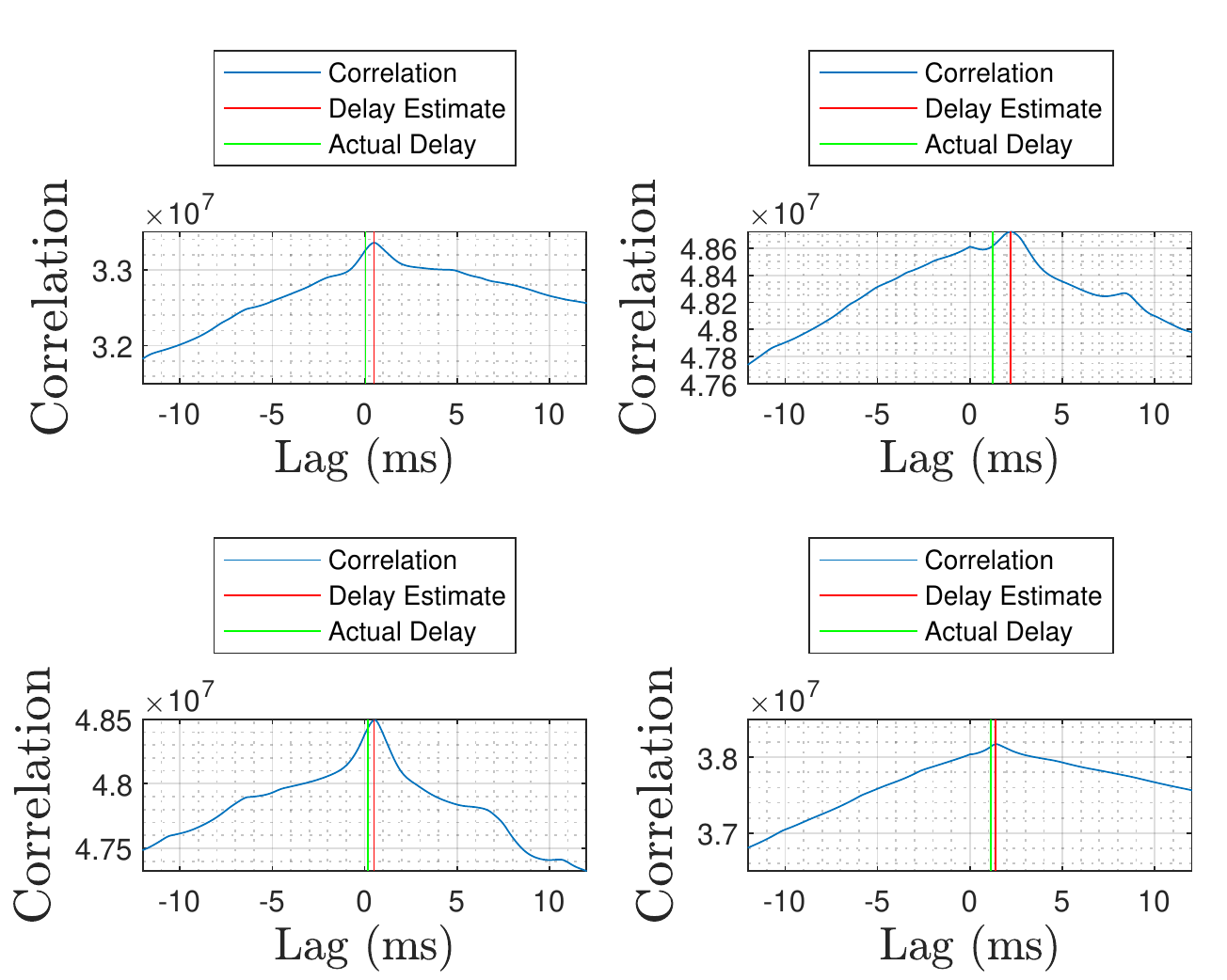}
    \caption{Comparison of 2-cell cross-correlation with delay estimate. Cross-correlation shown in blue, estimated delay as a vertical red line, actual delay as a vertical green line}
    \label{fig:sample2CellCorrPlots}
\end{figure}

Fig \ref{fig:sample2CellCorrPlots} shows a number of plots representing this delay estimation. Each subplot shows the cross-correlation of the 2 signals at a number of lag values from $-12ms$ to $+12ms$, along with the location of the estimated link delay (i.e. the location of the first positive peak) and the location of the actual link delay (as specified by the \emph{delay} parameter).

\begin{figure}[ht]
    \centering
    \includegraphics[width=0.48\textwidth]{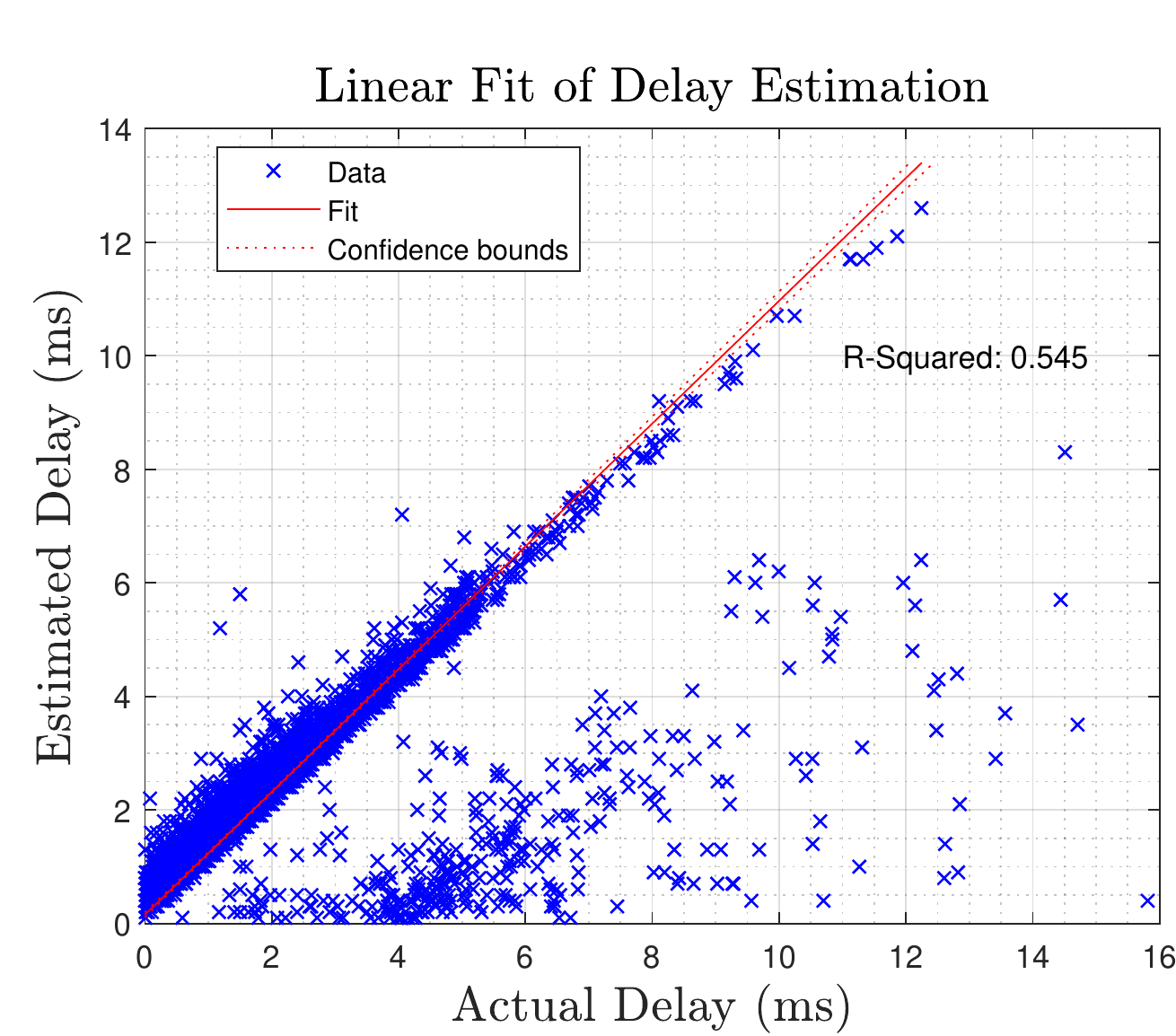}
    \caption{Correlation of delay estimate vs actual delay with least-squares linear model and associated R-squared score.}
    \label{fig:2CellDelayLFit}
\end{figure}

We applied the delay estimation method against all the simulated networks (N=10,000) and the correlation between the estimated delay and the actual delay was analysed. 
We fit the linear model to the data to determine the correlation between the estimation and the actual delay. A scatter-plot of the estimation and actual delay is shown in Fig. \ref{fig:2CellDelayLFit} along with the least-squares linear fit and associated R-squared score. The mean-squared error of this linear model was 1.0881ms. These results not only validate our delay estimator but demonstrate its advantage to existing morphological inference structures that are far more complex than the presented model. For example, neuron microcircuits would be comprised of tens of cells.

\subsubsection{Entropy and Mutual Information}
As discussed previously, the entropy of the cells is calculated by discretising the spike train and encoding each spike as a binary symbol. Using a symbol sequence it is then possible to apply the definition of mutual information in Eq. \ref{eq:mutualInf} to determine the entropy of the individual cells and the mutual information of the cell-to-cell connection. Here we apply the previously introduced delay estimation model that allows us to shift the spike train of the tail-cell such that the spike-response in that cell becomes virtually instantaneous. This is done to investigate whether or not adjusting the delay in the link will effect the mutual information calculation.

\begin{figure}[ht]
    \includegraphics[width=0.48\textwidth]{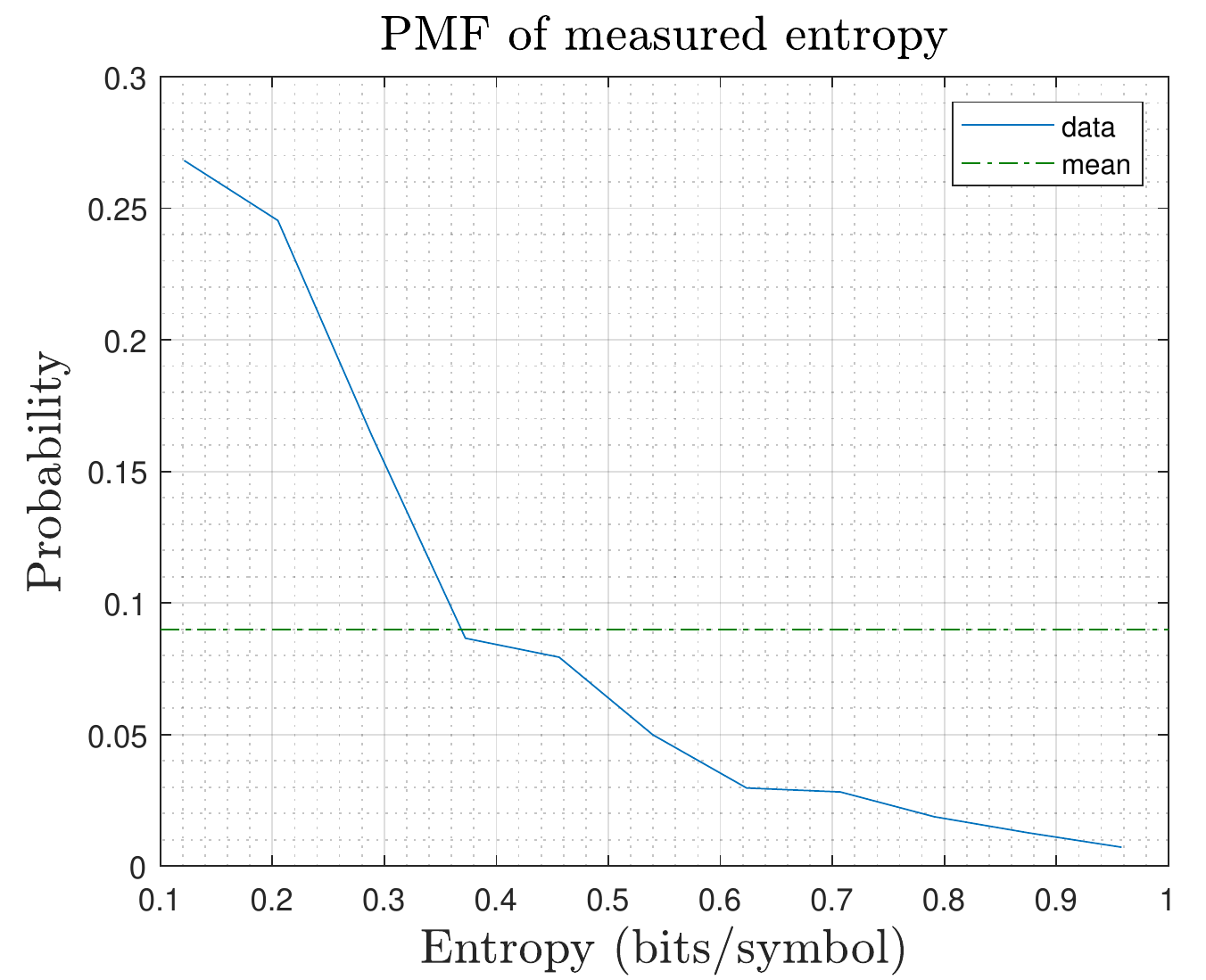}
    \caption{PMF of calculated single-cell entropy based on output spike trains}
    \label{fig:cellEntPmf}
\end{figure}

\begin{figure}[ht]
    \includegraphics[width=0.48\textwidth]{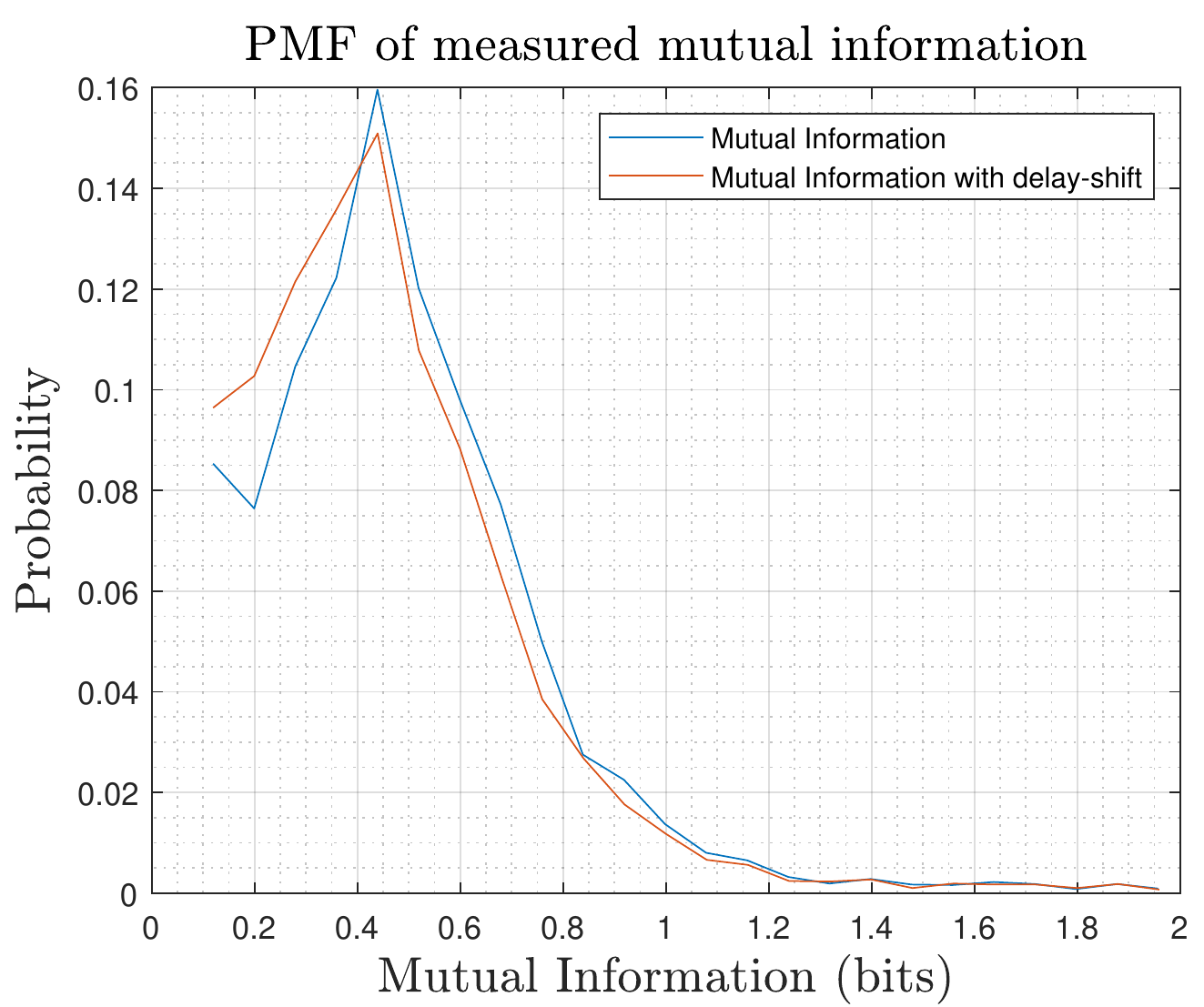}
    \caption{PMF plots of calculated mutual information for the measured spike-train (blue) and the delay-estimate shifted spike-train (orange)}
    \label{fig:mutualInfoPmf}
\end{figure}

The probability mass function (PMF) of the calculated single-cell entropy is shown in Fig. \ref{fig:cellEntPmf}. The majority of the cells have an entropy of below $0.4$ bits/symbol. The PMF of the calculated mutual information is shown in Fig. \ref{fig:mutualInfoPmf}. We can see that the mutual information of the networks is well-distributed, with a mean of around $0.5$ bits. The effect of shifting the spike-train based on the delay estimate has little to no effect on the distribution of the mutual information since the spike delay variability is low with the used cells. We conclude that these two metrics should be used in parallel for the training of the classification models since they measure different aspects of the communication of neurons.

\subsection{2-Cell Classification and Model Training}

In Fig. \ref{fig:sample2CellPlots} a number of spike-trains from different cell types were shown. Referring back to this, it is clear that different cells respond differently to the same stimulus, with the output spike train differing in intensity (frequency of spikes) as well as in the "settle-down" period after a spike was triggered (i.e. the fall-time response of the membrane voltage). It is these differences that we attempt to characterise through the use of the linear-filter portion of the linear-Nonlinear-Poisson (LNP) cascade model.

The first step in this process is to simulate some 2-cell networks again to generate test data. In this instance, the head-cell acts only as a stimulus generator, and we treat the soma-membrane potential of this cell as the "input" voltage to the linear system described in Eq.  (\ref{eq:firLinSys}). To limit the number of variables in this investigation, the type of cell used as the head-cell was constant across all simulations (layer 1, DAC m-type, bNAC e-type) while the tail-cell was treated as the cell-under-test and so was varied between simulations. As well as this, variations in link-level parameters (number of synapses etc.) were kept to a minimum to reduce the number of variables. A large number of networks were generated (N=30,000) to produce a dataset of sufficient size for classifier training.

\begin{figure}[ht]
    \centering
    \includegraphics[width=0.48\textwidth]{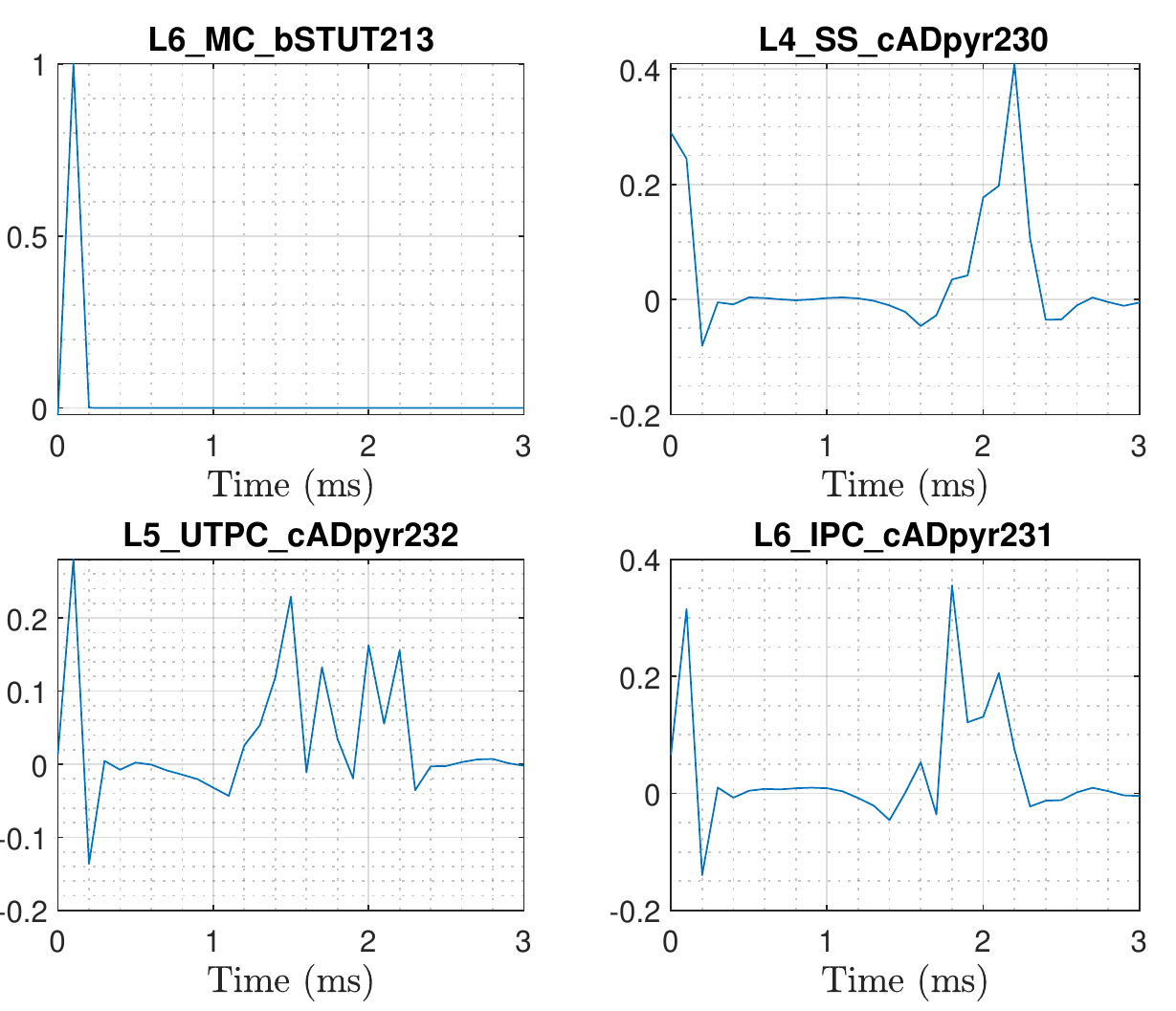}
    \caption{Sample impulse-response characterisation of different cells.}
    \label{fig:sampImpRes}
\end{figure}

Following simulation and before training the classifiers, the filter estimation step described in Eq. (\ref{eq:estFilter}) was applied to extract the filter coefficients as features from the cell-response data. Fig. \ref{fig:sampImpRes} shows the filter coefficients (FIR impulse response) of 4 different cells, with a filter-order of 64. It is clear from this diagram that the impulse-response estimation of the neuronal cell is capable of differentiating between the cell types. Using these estimated filter coefficients, we are now able to begin training the classifiers.

Using the \emph{Classification Learner} tool in Matlab, we can train and compare a set of different classifiers, inspecting the accuracy, confusion matrix, and receiver-operating characteristic (ROC) curve for each. The first classifier to investigate is the layer predictor, which takes the estimated filter coefficients as input, and attempts to estimate what layer the observed cell belongs to. We use this tool to train and analyse an SVM and decision tree classification systems, while we use \emph{RapidMiner} to train the random forest and artificial neural network classifiers. 

The performance of the classifiers is shown in Tab. \ref{tbl:layerClassifierPerf}, \ref{tbl:mtypeClassifierPerf} and \ref{tbl:etypeClassifierPerf}. In each table, we compare the performance of the different classification algorithms (Decision Tree, SVM etc.) in the classification of the separate cell sub-group types (layer, m-type, e-type). For each classifier, we state the corresponding accuracy. The accuracy is calculated as the ratio of correct estimations versus the total number of estimations. We also state the "factor improvement" of the classifier. This is calculated as the factor by which the classifier improves over the equivalent accuracy of random guessing in the classification space. For example, with 5 layer classes, the equivalent "random guess" accuracy is 20\%, and so a trained classifier with an accuracy of 40\% would have a factor improvement of 2. 

Table \ref{tbl:layerClassifierPerf} shows the performance of the classifiers in the prediction of cell layer-type, Table \ref{tbl:mtypeClassifierPerf} for m-type prediction, and Table \ref{tbl:etypeClassifierPerf} for e-type prediction. In every case, the SVM classifier has the highest accuracy, with the Neural Network being slightly behind. Of the tree-based classifiers, interestingly the decision tree based classifier has a constant higher accuracy than the random forest classifier, despite the latter being a variant of the former. That is based on the construction of the trees used in both and the simple data structures used for training, which provided the construction of efficient classifications trees for the tree-based classifiers.

\newcolumntype{L}[1]{>{\raggedright\let\newline\\\arraybackslash\hspace{0pt}}m{#1}}
\newcolumntype{C}[1]{>{\centering\let\newline\\\arraybackslash\hspace{0pt}}m{#1}}
\newcolumntype{R}[1]{>{\raggedleft\let\newline\\\arraybackslash\hspace{0pt}}m{#1}}

\begin{table}[h]
    \centering
    \caption{Performance of different algorithms at classifying cell-layer (5 classes, 20\% equivalent random guessing accuracy)}
    \begin{tabular}{|C{1.5cm}||c|c|c|c|}
        \hline
        Classifier & Decision Tree & Random Forest & SVM & NN\\
        \hline\hline
        Accuracy & 52.2\% & 41.43\% & 62.5\% & 60.98\%\\
        \hline
        Factor\newline Improvement & 2.61 & 2.07 & 3.13 & 3.05 \\
        \hline
    \end{tabular}
    \label{tbl:layerClassifierPerf}
\end{table}

\begin{table}[h]
    \centering
    \caption{Performance of different algorithms at classifying cell m-type (25 classes, 4\% equivalent random guessing accuracy)}
    \begin{tabular}{|C{1.5cm}||c|c|c|c|}
        \hline
        Classifier & Decision Tree & Random Forest & SVM & NN\\
        \hline\hline
        Accuracy & 42.6\% & 35.07\% & 63.7\% &57.99\%\\
        \hline
        Factor\newline Improvement & 10.65 & 8.77 & 15.93 &13.92\\
        \hline
    \end{tabular}
    \label{tbl:mtypeClassifierPerf}
\end{table}

\begin{table}[h]
    \centering
    \caption{Performance of different algorithms at classifying cell e-type (14 classes, 7.143\% equivalent random guessing accuracy)}
    \begin{tabular}{|C{1.5cm}||c|c|c|c|}
        \hline
        Classifier & Decision Tree & Random Forest & SVM & NN\\
        \hline\hline
        Accuracy & 63.8\% & 54.47\% & 75.3\% &73.94\%\\
        \hline
        Factor\newline Improvement & 8.93 & 7.63 & 10.54 &10.35\\
        \hline
    \end{tabular}
    \label{tbl:etypeClassifierPerf}
\end{table}

Another set of metrics that are used to quantify the performance of a classifier are the class precision and the class recall. The class recall is calculated per-class as the ratio of correct predictions of a class versus the number of observations of that class. The class precision is again calculated per-class as the ratio of correct predictions of a class versus the overall number of estimations per class. As these metrics identify the performance of the model on a class-by-class basis, they result in a large number of individual values to compare. When we consider that we are comparing three estimator groups with a relatively high number of classes per group (five for layer estimator, 25 for m-type, and 14 for e-type), as well as the fact that each estimator group must be compared against four different classification algorithms, it becomes difficult to critically compare the performance of each individual class. In general, however, when analysing such metrics from the single classifier, a \emph{confusion matrix} is used. The confusion matrix of the SVM-based layer estimator is shown in Table \ref{fig:svmConfMatLayer}. The confusion matrix tabulates the estimations of a given classifier as a grid of "True Class" versus "Predicted Class". In this way, it can show how often an observation of a given class is estimated as belonging to some other class. As such, the diagonal (shown as green in the confusion matrix figure) shows the proportion of the correct predictions in the validation set (where true class equals predicted class). The right-hand bar shows the proportional difference of the \emph{true positive rate} and the \emph{false negative rate}. The true-positive rate is equivalent to the class recall.
While the confusion matrices for every single classifier investigated in this study are not included, the relative proportions between them tend to be similar (i.e. very high prediction accuracy between layer 1 and layer 6, with reduced accuracy in the intermediate layers).

\begin{figure*}[ht]
    \centering
    \includegraphics[width=0.9\textwidth]{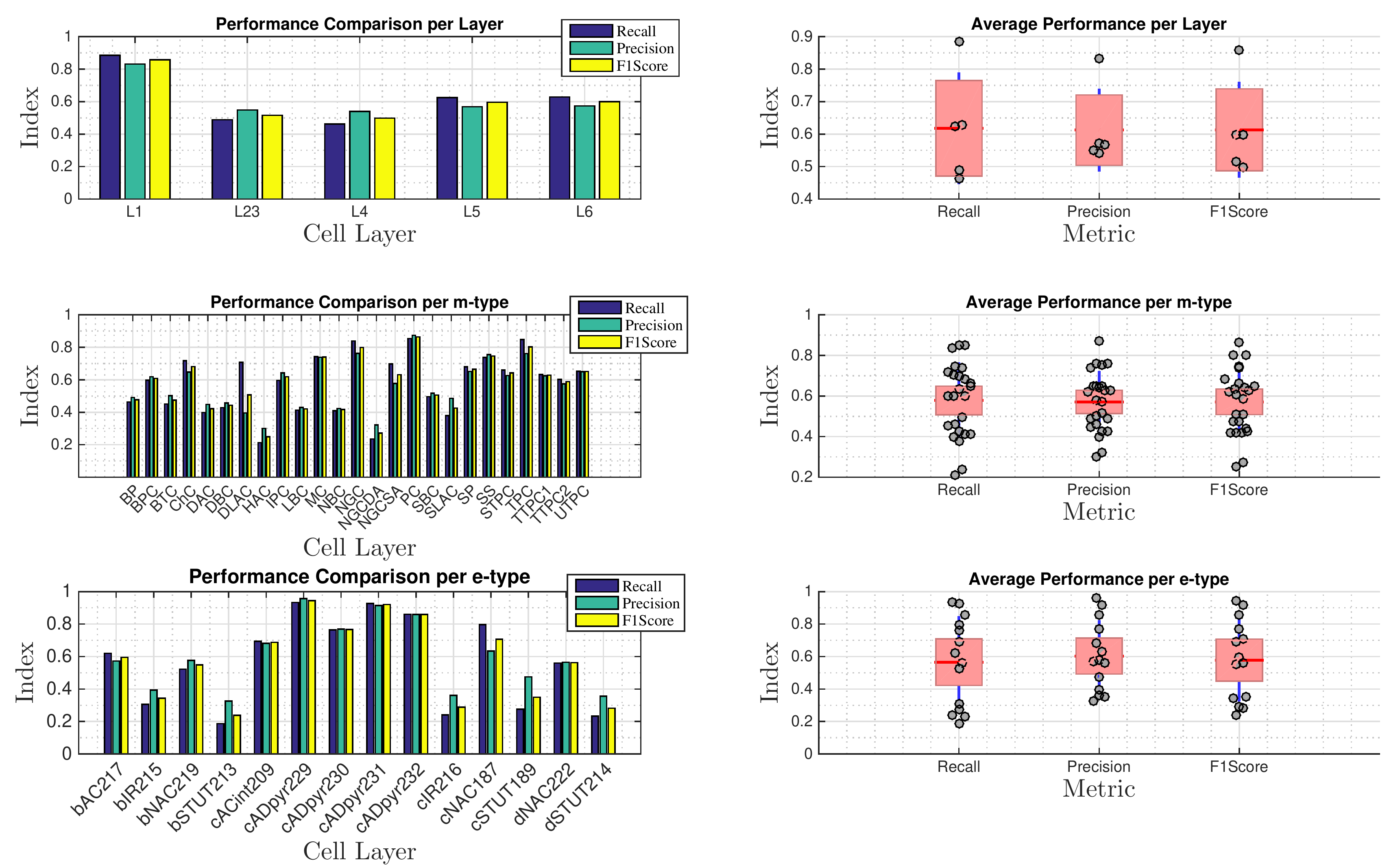}
    \caption{Results for the comparison of different types of cell layers, m-types and e-types considering recall, precision and F1score performance.}
    \label{fig:comparison_types}
\end{figure*}

Performance comparison between layers, m-type as well as e-type present also variability for metrics such as recall, precision and f1score. In Fig \ref{fig:comparison_types}, we show for an average close to 0.6 of all metrics for all layers, we can observe that classifying cells in layer one is often easier. Many reason can be made for that, but it all converges in the easy propagation of spike signals in that layer. The variability in the performance is higher when considering the comparison of m-types and e-types. Our results show that the link in cell structure and properties present different levels of communication that can be detected. While in some cell types this detection is still poor (like $NGCDA$ m-type or $bSTUT213$ e-type) in other cells present really high classification performance (like $PC$ m-type or $cADpyr229$ e-type). The variability in the presented results can be attributed to the spike variability for cell-cell communication as well as fine-tuning of the machine learning techniques used. A future solution to be verifiable for this problem is the not only adjusting the machine learning configuration for each type of cell, but also look at different types of machine learning technique and their combination. Even though this seems an ardours tasks, investigating these proposed solutions will bring tremendous benefits in detecting novel patterns of communication as well as structure in the brain that hasn't been identified before.

\begin{table}[ht]
    \centering
    \caption{Confusion Matrix of SVM layer-classifier}
\begin{tabular}{cc|c|c|c|c|c|c|c|c|}
\cline{3-7} \cline{9-10}
                                                   &      & \multicolumn{5}{c|}{Predicted Class}                                                                                                                                         &  &                                                                       &                                                                        \\ \cline{3-7}
                                                   &      & L1                           & L2/3                         & L4                           & L5                                     & L6                                     &  & \multirow{-2}{*}{\begin{tabular}[c]{@{}c@{}}True\\ Pos.\end{tabular}} & \multirow{-2}{*}{\begin{tabular}[c]{@{}c@{}}False\\ Neg.\end{tabular}} \\ \cline{1-7} \cline{9-10} 
\multicolumn{1}{|c|}{}                             & L1   & \cellcolor[HTML]{009901}90\% & \cellcolor[HTML]{FF9A97}8\%  & \cellcolor[HTML]{FFCCC9}2\%  & \cellcolor[HTML]{FFCCC9}\textless{}1\% & \cellcolor[HTML]{FFCCC9}\textless{}1\% &  & \cellcolor[HTML]{009901}90\%                                          & \cellcolor[HTML]{FF9A97}10\%                                           \\ \cline{2-7} \cline{9-10} 
\multicolumn{1}{|c|}{}                             & L2/3 & \cellcolor[HTML]{FF9A97}13\% & \cellcolor[HTML]{34FF34}48\% & \cellcolor[HTML]{FF9A97}11\% & \cellcolor[HTML]{FF9A97}11\%           & \cellcolor[HTML]{FF9A97}16\%           &  & \cellcolor[HTML]{34FF34}48\%                                          & \cellcolor[HTML]{FF4C4C}52\%                                           \\ \cline{2-7} \cline{9-10} 
\multicolumn{1}{|c|}{}                             & L4   & \cellcolor[HTML]{FFCCC9}3\%  & \cellcolor[HTML]{FF9A97}14\% & \cellcolor[HTML]{34FF34}47\% & \cellcolor[HTML]{FF9A97}21\%           & \cellcolor[HTML]{FF9A97}15\%           &  & \cellcolor[HTML]{34FF34}47\%                                          & \cellcolor[HTML]{FF4C4C}53\%                                           \\ \cline{2-7} \cline{9-10} 
\multicolumn{1}{|c|}{}                             & L5   & \cellcolor[HTML]{FFCCC9}1\%  & \cellcolor[HTML]{FF9A97}11\% & \cellcolor[HTML]{FF9A97}10\% & \cellcolor[HTML]{32CB00}62\%           & \cellcolor[HTML]{FF9A97}16\%           &  & \cellcolor[HTML]{32CB00}62\%                                          & \cellcolor[HTML]{FD6864}38\%                                           \\ \cline{2-7} \cline{9-10} 
\multicolumn{1}{|c|}{\multirow{-5}{*}{\rotatebox{90}{True Class}}} & L6   & \cellcolor[HTML]{FFCCC9}1\%  & \cellcolor[HTML]{FF9A97}10\% & \cellcolor[HTML]{FF9A97}11\% & \cellcolor[HTML]{FF9A97}12\%           & \cellcolor[HTML]{32CB00}65\%           &  & \cellcolor[HTML]{32CB00}65\%                                          & \cellcolor[HTML]{FD6864}35\%                                           \\ \cline{1-7} \cline{9-10} 
\end{tabular}
    \label{fig:svmConfMatLayer}
\end{table}

\subsubsection{Network Tomography for Cellular Classification}

As the SVM classifier had the best performance in comparison to the other classification models, this is the model that was applied in the reconstruction of the 4-leaf star networks. The process here was similar to the previous experiments: several unique networks were generated, simulated, and their data-points measured and analysed. In this case, the networks generated were of the 4-leaf star topology previously discussed. The central node was constrained to be of the same cell type used in the training of the models (layer 1, DAC m-type, bNAC e-type), while the star cells were varied. For each measurement from the soma-membrane of a star node, the characteristic filter was estimated using the same process discussed previously, and the filter coefficients were passed through the pre-trained classifiers.

The performance results of the topology reconstruction are shown in Tab. \ref{tbl:wholeClassifierPerf}. Here, we tabulate the accuracy of the classifier chain in estimating layer, m-type, and e-type groups, as well as the "whole-cell" classification accuracy. We define whole-cell accuracy as the cases where all three sub-groups were correctly estimated. The factor-improvement for the whole-cell prediction is significantly higher than for any other group, while the sub-group accuracy is quite similar to those in the 2-cell networks.

\begin{figure*}[ht]
    \centering
    \includegraphics[width=0.7\textwidth]{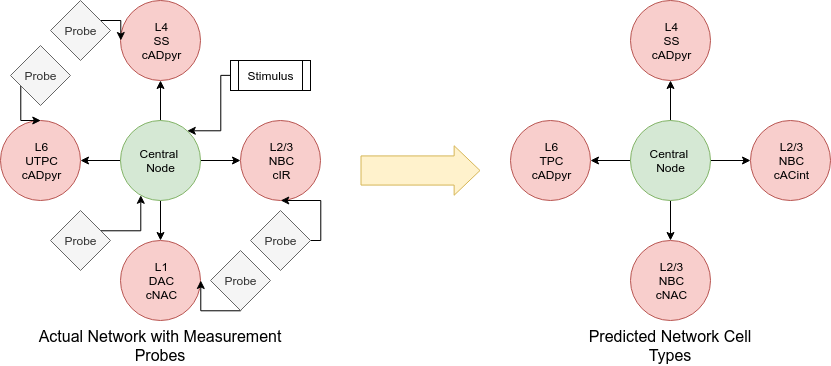}
    \caption{Sample reconstruction of network}
    \label{fig:4CellRecon}
\end{figure*}

Fig. \ref{fig:4CellRecon} shows a sample reconstruction of a 4-leaf network from probe measurements. The left side of the figure represents the network as it was simulated, with probes at the network endpoints, probe in the central node, and a stimulus on the central node, along with the actual cell types of the leaf nodes. The right side of the figure shows the reconstructed topology based on the cell-type estimations from the SVM classifier chain. The left side is the original network topology, and the right side is the predicted network topology.

\begin{table}[ht]
    \centering
    \caption{Performance of 4-leaf star topology reconstruction}
    \begin{tabular}{|C{1.9cm}||c|c|c|c|}
        \hline
        Prediction & Layer & m-type & e-type & whole-cell \\
        \hline\hline
        \# classes & 5 & 25 & 14 & 1750\\
        \hline
        Equiv. random guess accuracy & 20.0\% & 4.0\% & 7.143\% & 0.0571\%\\
        \hline
        Classifier\newline accuracy & 61.82\% & 56.34\% & 64.62\% & 36.23\%\\
        \hline
        Factor\newline Improvement & 3.091 & 14.085 & 9.05 & 634.5\\
        \hline
    \end{tabular}
    \label{tbl:wholeClassifierPerf}
\end{table}

\section{Discussion}
\label{sec:discFuture}

The presented objective classification of cells from neural communication data shows promising results, with relatively high accuracy considering the high number of classes to which the models were estimating. The SVM-based classifier had the highest level of accuracy, with the artificial neural network being close behind in performance. This was surprising, as neural networks are widely used for high-order classification problems. It is possible, however, that the close level of accuracy between the two is indicative of the limit of the features on which we are predicting and that the level of information that the features hold does not allow for accuracy above this level. This is supported by the fact that tweaks to the classifiers' parameters (i.e. kernel size in SVMs and hidden-layer size in the neural networks) did not have any noticeable effect on the accuracy, indicating that they are close to maximising the classification accuracy based on these features. Future work in this area could, therefore, look at the extension or re-engineering of the feature-space. 
Another notable analysis was the size of the linear filter used as the characteristic features (the order of the FIR filter). Several different filter sizes were investigated, however, we found that above about 64 coefficients, the accuracy did not improve. Here we can see that the majority of the characterisation between cell types is in the central coefficients, with reduced variance as we move away from the 0-point. Increasing the order of the filter will only add detail to the two extremes of this impulse response, which adds features that seem to bear little characterising information.

The classifier with the highest performance in all cases was the SVM-based model. In classifying the layer type, the SVM model was capable of achieving 62.5\% accuracy, a factor improvement of 3.13 over random guessing. By looking at the confusion matrix for this classifier in Figure \ref{fig:svmConfMatLayer}, we can see that the classifier has very good performance in classifying between layer 1 and layer 6 with close to 100\% accuracy between these classes, where only 1\% of layer 6 cells were predicted as layer 1, and less than 1\% of layer 1 cells were predicted as layer 6. The overall accuracy decreases as the intermediate layers are added, with the worst-performing component being incorrectly predicting a layer 4 cell as layer 5 for 21\% of the time. It is evident, therefore, that the highest degree of separation using the FIR-filter estimation is between layer 1 and layer 6.
In the reconstruction of the 4-leaf topology through endpoint measurements using the trained SVM classifier, the individual sub-group accuracy values tend to reflect those of the 2-cell networks with a layer-estimation accuracy of 61.82\%, an m-type accuracy of 56.34\%, and an e-type accuracy of 64.62\% representing a factor improvement of 3.09, 14.09, and 9.05 respectively. The interesting result in this investigation, however, is the whole-cell estimation (where each of the individual sub-group estimators was correct) with an overall accuracy of 36.23\% and a huge factor improvement of 634.5. 

Several areas could be worked on in future to extend the presented investigation. 
One of them is the variation of the central/stimulus cell. Throughout the production of our simulation database, the presynaptic cell was kept constant to minimise variables. In practice, however, many different cell types may interconnect and so the ability to classify a cell regardless of the presynaptic cell type is important. This can create a large set of dynamics that must be also captured by the classification model \cite{ofer2019axonal}. Conceptually, the performance should be relatively similar to the features used in the classifiers are based on the impulse response of the postsynaptic cell, rather than the characteristics of the presynaptic cell. Another consideration for future work would be to assume multi-path networks enabling classification and inference of higher complexity cortical structures. In this investigation, we dealt solely with a single path (one cell to one cell) connections. In practice, a given cell may be stimulated by a number of presynaptic cells. This can be characterised using the LNP model by using multiple linear filters and computing the output based on a combination of the individual coefficients \cite{lnp}. Similarly, in our study, we used only excitatory synaptic connections to generate measurements with a large number of characterising spikes. Even, metrics that rely on molecular information can also be added \cite{barros2018feed}, which can potentially classify cell types over time and detecting as well as predicting disease states in the tissue \cite{barros2018multi}. Future work could, therefore, look at the classification of cells based on various combinations of excitatory to inhibitory connections, which can alter the training dataset configuration and require new approaches for removing dataset bias.

\section{Conclusion}
\label{sec:conc}

To contribute to advancements in objective morphological classification of neurons, we present a technique that uses information network tomography to classify and infer the neurons' morphological types and networks in the somatosensory cortex of the brain. We use the information transfer metrics as well as extend existing signal processing tools to accommodate larger cortical structures in our prediction analysis, such as small-world topologies.
We have shown that the classification of neuronal cell types and networks is possible through the characterisation of the cell's input-output impulse response. With an average accuracy of around 58\% across the layer, m-type, and e-type class groups, the SVM-based classifier outperforms decision tree, random forest, and even artificial neural network classifiers. We have also shown that the trained classifier can then be applied in the reconstruction of the cell-types in various forms of the 4-leaf star topology, estimating the cell-type of each leaf node to a promising degree of accuracy. This was only possible with the development of two simulation support libraries for the NEURON simulator, Neurpy and NeurGen. We used these libraries to create 10,000 network topologies that were part of the classifier's training and accuracy analysis, also recall, precision and F1score of five layers, 25 cell m-types, and 14 cell e-types. Besides the promising performance of our proposed system, it is clear that more research is required to improve the robustness of this form of classification to any usable degree. While the investigations carried out in this study were preliminary regarding the extremely wide scope of neurology, molecular communication, network tomography, and information theory, it is evident that the individual topics covered may be used in future for the robust and reliable in-vivo detailed characterisation of neuronal structures using future miniaturised implantable devices.

\addtolength{\textheight}{-12cm}   

\bibliography{refs.bib}
\bibliographystyle{ieeetr}

\end{document}